\documentclass{PoS}

\usepackage{amsfonts,amsmath,amssymb,bm}
\usepackage{graphicx} 

\newcommand{\be}{\begin{equation}}
\newcommand{\bea}{\begin{eqnarray}}
\newcommand{\ee}{\end{equation}}
\newcommand{\eea}{\end{eqnarray}}
\newcommand{\sla}{\slash \hspace{-0.22cm}}

\def\quark{\widehat{X}}

\def\G{\Gamma}
\def\s#1{{\scriptscriptstyle #1}}

\def\noeq#1{(\ref{#1})}
\def\1eq#1{Eq.~(\ref{#1})}

\def\2eqs#1#2{Eqs.~(\ref{#1}) and~(\ref{#2})}
\def\3eqs#1#2#3{Eqs.~(\ref{#1}),~(\ref{#2}) and~(\ref{#3})}

\def\fig#1{Fig.~\ref{#1}}

\def\DeltaQ{\Delta_\s{\nf}}
\def\mQ{m^2_\s{\nf}}

\def\kint{\int_k\!}
\def\diff#1{{\rm d}^#1}

\def\ie{{\it i.e.}, }

\def\n#1{({\it #1}\,)}

\def\nf{N_{\!f}}

\def\Y{Y}

\def\quark{\widehat{X}}

\def\pslash{p\hspace{-0.18cm}\slash} 
\def\qm{{\cal M}}

\title{Gluon propagator with dynamical quarks}

\ShortTitle{Gluon propagator with dynamical quarks}

\author{\speaker{Joannis Papavassiliou}\\
        Department of Theoretical Physics and IFIC,\\ 
        University of Valencia and CSIC,\\
        E-46100, Valencia, Spain.\\
        E-mail: \email{Joannis.Papavassiliou@uv.es}}

\abstract{ We review recent work on  the effects of quark loops on the
 gluon  propagator  in  the   Landau  gauge,  relying  mainly  on  the
 Schwinger-Dyson equations  that describe the  two-point sector of QCD.
 Particularly important in  this context is the detailed  study of how
 the standard  gluon mass  generation mechanism, which  is responsible
 for  the infrared  finiteness of  the quenched  gluon  propagator, is
 affected  by  the inclusions  of  dynamical  quarks.   This issue  is
 especially relevant  and timely,  given the qualitative  picture that
 emerges  from  recent unquenched  lattice  simulations.  Our  results
 demonstrate clearly that the gluon mass generation persists, and that
 the   corresponding  saturation  points   of  the   unquenched  gluon
 propagators  are progressively  suppressed,  as the  number of  quark
 flavors increases.}

\FullConference{QCD-TNT-III-From quarks and gluons to hadronic matter: A bridge too far?,\\
		2-6 September, 2013\\
		European Centre for Theoretical Studies in Nuclear Physics and Related Areas (ECT*), Villazzano, Trento (Italy)}

\begin{document}

\section{Introduction}

In recent years outstanding 
progress has been made in our understanding 
of various aspects of the nonperturbative dynamics of Yang-Mills theories,
through the fruitful combination of a variety 
of approaches and techniques
~\cite{Aguilar:2006gr,Aguilar:2008xm,Boucaud:2008ky,Alkofer:2000wg,Huber:2012kd,
Cucchieri:2007md,Bogolubsky:2009dc,Pawlowski:2005xe,Szczepaniak:2011bs,Campagnari:2011bk,Quandt:2013wna}. 
In fact, we appear to have obtained 
a rather firm grasp on the infrared behavior of the fundamental 
Green's (correlation) functions of QCD, such as gluon, ghost, and quark 
propagators, as well as some of the basic vertices of the theory. 
Of course, the gluon propagator occupies deservedly a prominent position in this ongoing endeavor,  
because its infrared behavior is intimately connected 
to the fundamental question of if and how Yang-Mills theories 
generate dynamically a mass scale.

The full gluon propagator in the Landau gauge assumes the general form 
\be
i{\Delta}_{\mu\nu}(q)=-i{\Delta}(q^2)P_{\mu\nu}(q); \qquad 
P_{\mu\nu}(q)=g_{\mu\nu}-{q_\mu q_\nu}/{q^2},
\label{prop-def}
\ee 
where $\Delta(q^2)$ is related to the scalar form factor of the  
gluon self-energy $\Pi_{\mu\nu}(q)=\Pi(q^2)P_{\mu\nu}(q)$  through
\be
\Delta^{-1}(q^2)=q^2+i\Pi(q^2) \,.
\ee

As has been demonstrated in detail~\cite{Aguilar:2011xe,Binosi:2012sj}  within the general 
framework of the Schwinger-Dyson equations (SDEs)~\cite{Roberts:1994dr}, 
the nonperturbative dynamics of pure Yang-Mills theories gives rise to a dynamical (momentum-dependent) 
gluon mass~\cite{Cornwall:1981zr,Philipsen:2001ip,Nair:2013kva}, 
which accounts for the infrared finiteness of the quenched $\Delta(q^2)$, 
observed in large-volume lattice simulations, both in  
$SU(2)$~\cite{Cucchieri:2007md} and in $SU(3)$~\cite{Bogolubsky:2009dc}.
In particular, in Minkowski space, 
the gluon propagator may be described in terms of two basic functions,  $J(q^2)$ and $m^2(q^2)$
\begin{equation}
\label{massiveprop}
\Delta^{-1}(q^2) = q^2 J(q^2) - m^2(q^2), 
\end{equation}
where, in the limit $q^2\to 0$, we have that  $q^2 J(q^2) \to 0$, whereas $m^2(0)\neq 0$. 
This property of the mass function $m^2(q^2)$ accounts for the fact that $\Delta(q^2)$ 
saturates at a non-vanishing constant value in the deep infrared~\cite{Aguilar:2008xm,Cornwall:1981zr}. 
The presence of this effective mass explains,
in addition, the finiteness of the dressing function, $F(q^2)$~\cite{Aguilar:2008xm,Boucaud:2008ky}, which is 
related to the full ghost propagator, $D(q^2)$, by 
\begin{equation}
D(q^2) = \frac{F(q^2)}{q^2}\,.
\end{equation}

It is interesting to point out that, in $d=4$,  
 the function  $J(q^2)$ diverges logarithmically in the deep infrared, 
due to the fact that the ghost-loop contributing to it contains massless ghost propagators~\cite{Aguilar:2013vaa}. 

It turns out that  
new lattice simulations involving dynamical quarks~\cite{Ayala:2012pb} 
furnish further valuable information that permits us to scrutinize in much more detail
the general dynamical scenario described above. 
In particular, the unquenched gluon propagators 
continue to saturate in the infrared, which  
suggests that 
the mass generation mechanism persists 
in the presence of quark loops. In fact, the observed considerable 
suppression of the value of their 
saturation points compared to the 
quenched ones clearly suggests that 
the corresponding gluon masses increase. 

The purpose of this presentation is to report on recent work in the continuum,  
which fully confirms the general trends displayed by the lattice results of~\cite{Ayala:2012pb}.  
This particular study has been carried out 
within the framework provided by the synthesis of the 
pinch technique (PT)~\cite{Cornwall:1981zr,Cornwall:1989gv,Pilaftsis:1996fh,Binosi:2009qm} 
with the background field method (BFM)~\cite{Abbott:1980hw}, known in the 
literature as the PT-BFM scheme~\cite{Aguilar:2006gr}. 
For a related analysis, see~\cite{Bashir:2014fba}.
Note also that in the context of the so-called ``scaling'' solutions~\cite{Alkofer:2000wg}
the unquenching effects have been considered in~\cite{Fischer:2003rp,Fischer:2005en}. 

The general philosophy underlying the work of \cite{Aguilar:2013hoa} may be summarized as follows.
 The momentum evolution of $m^2(q^2)$
is described by a homogeneous integral equation, to be referred to as the \textit{mass equation},
whose kernel depends on $\Delta(q^2)$ 
in a complicated way. 
The detailed numerical study of this particular equation, 
for pure $SU(3)$ Yang-Mills, revealed that its   
solutions depend strongly on the  
precise shape of $\Delta(q^2)$ through a wide range of momenta~\cite{Binosi:2012sj}. 
Of course, 
the inclusion of dynamical quarks modifies the form of the 
gluon propagator, to be denoted by $\DeltaQ(q^2)$; the modifications 
depend, among other things, on the number of quark families, $\nf$, 
and the values of the corresponding quark masses.
In the context of the SDEs, 
the behavior of 
$\DeltaQ(q^2)$ in the range of intermediate momenta 
may be obtained from the quenched gluon propagator $\Delta(q^2)$ 
by means of an approximate procedure, 
which attributes the main bulk of the ``unquenching'' to the {\it fully dressed} quark-loop graph, 
while higher loop contributions are considered to be subleading~\cite{Aguilar:2012rz}. The value of 
$\DeltaQ(0)$, however, is determined only in conjunction with the gluon mass equation,
in whose kernel one must implement the change $\Delta(q^2) \to \DeltaQ(q^2)$.
Thus, the study presented here finally boils down to the simultaneous solution of 
the mass equation and the master formula that controls the amount by which $\DeltaQ(q^2)$
deviates from the quenched $\Delta(q^2)$; for the latter we will use directly the lattice data of~\cite{Bogolubsky:2009dc}.


\section{Gluon mass generation in a nutshell.}
 
The gauge invariant generation of a gluon mass~\cite{Cornwall:1981zr} 
proceeds through the implementation 
of the  Schwinger mechanism~\cite{Schwinger:1962tn,Schwinger:1962tp}, which 
requires the existence of a very special type of 
nonperturbative vertices~\cite{Jackiw:1973tr,Cornwall:1973ts,Eichten:1974et,Poggio:1974qs,Ibanez:2012zk}, 
which within the PT-BFM framework are generically denoted by $\widetilde V$.
In particular:  

\n{i} The $\widetilde V$ vertices 
are {\it  longitudinally coupled}, and 
contain {\it massless poles}, which make possible that the SDE of the gluon propagator yields $\Delta^{-1}(0) \neq 0$.

\n{ii}  The aforementioned poles 
have nonperturbative origin:  they are {\it colored} composite states  with vanishing mass.  
They act as Nambu-Goldstone bosons, maintaining gauge invariance, 
but, are not associated with the spontaneous breaking of any continuous symmetry. 
In particular, their presence 
guarantees that the Ward identities and the Slavnov-Taylor identities 
of the theory remain {\it intact}, \ie they have the 
same form before and after mass generation. 

\n{iii} These  longitudinally coupled states {\it decouple} from on-shell amplitudes, 
and, in general, from physical observables.

\begin{figure}[!t]
\begin{center}
\includegraphics[scale=.95]{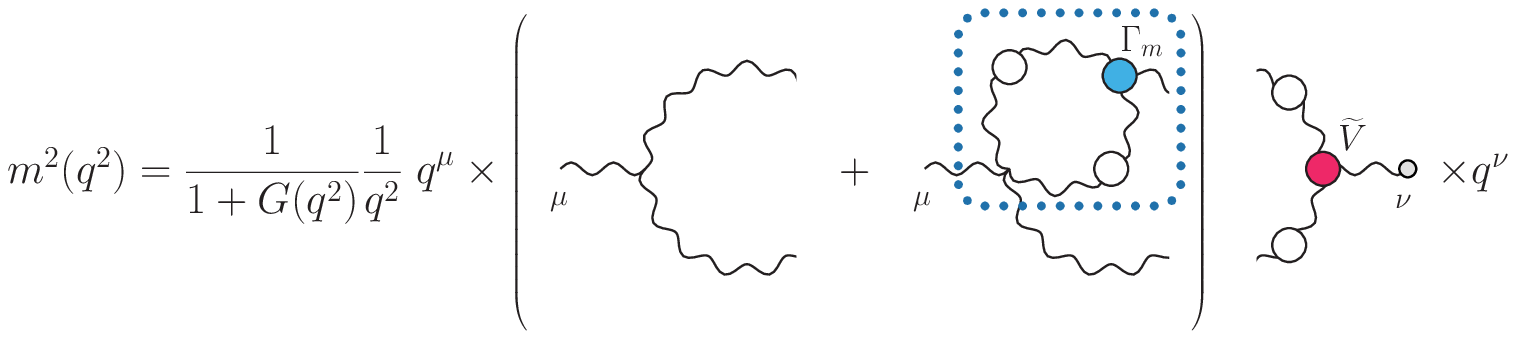}
\end{center}
\vspace{-0.5cm}
\caption{\label{meq-fig} The effective SDE satisfied by the dynamical gluon mass. The blue circle denotes  
the (conventional) fully-dressed three-gluon vertex $\Gamma_{\sigma\rho\beta}$, 
while the (red) $\widetilde{V}$ indicates a pole vertex 
whose external leg (with the little circled attached to it) is a background gluon.
Finally,  the perforated box marks the $Y(k^2)$, which represents the purely two-loop dressed correction to the one-loop dressed mass equation kernel.}
\end{figure}

The exact application of these notions at the level 
of the gluon propagator SDE is particularly subtle,
and has been discussed in great detail in the recent literature~\cite{Aguilar:2011xe,Binosi:2012sj}. 
The final upshot is the derivation of a homogeneous integral equation, valid only in the Landau gauge,  
which controls the momentum evolution of the gluon mass. 
It is given by~\cite{Binosi:2012sj} (see also~\fig{meq-fig}) 
\be
m^2(q^2) = -\frac{g^2 C_A}{1+G(q^2)}\frac{1}{q^2}\int_k m^2(k^2) \Delta_\rho^\mu(k)\Delta^{\nu\rho}(k+q)
{\cal K}_{\mu\nu}(k,q),
\label{masseq}
\ee
with 
\bea
{\cal K}_{\mu\nu}(k,q) &=& [(k+q)^2 - k^2] \left\{ 1 - [\Y(k+q) + \Y(k)]\right\}g_{\mu\nu}
\nonumber \\
&+& [\Y(k+q)-\Y(k)](q^2 g_{\mu\nu}-2q_\mu q_\nu). 
\label{massK}
\eea
The quantity $Y$ corresponds to the subdiagram nested inside the 
two-loop dressed graph of \fig{meq-fig},  given by 
\be
\Y(k^2)=\frac{g^2 C_A}{4 k^2} \,k_\alpha\! \int_\ell\!\Delta^{\alpha\rho}(\ell)
\Delta^{\beta\sigma}(\ell+k)\Gamma_{\sigma\rho\beta}(-\ell-k,\ell,k),
\label{theY}
\ee
with $\Gamma_{\sigma\rho\beta}$  the full three-gluon vertex, 
and $C_A$ the Casimir eigenvalue in the adjoint representation [$C_A=N$ for $SU(N)$].
In the  above equations we used the short-hand notation 
\mbox{$\int_k=\mu^\epsilon\int\!\diff{d}k/(2\pi)^d$} to denote the dimensional regularization measure, where $d=4-\epsilon$ 
is the space-time dimension and $\mu$ the 't Hooft mass. 

Finally, the function $G(q^2)$ corresponds to the 
$g_{\mu\nu}$ component of a special two point-function, which constitutes a 
key ingredient in a set of 
powerful identities, relating the conventional Green's functions 
to those of the BFM~\cite{Binosi:2009qm,Grassi:1999tp}. In particular, 
for the case of the conventional gluon propagator, $\Delta$, and the PT-BFM gluon propagator, denoted 
in the literature by $\widehat\Delta$, 
the corresponding identity reads 
\begin{equation}\label{propBQI}
\Delta(q^2) = [1+G(q^2)]^2 \widehat{\Delta}(q^2);
\end{equation}
its application at the level of the gluon SDE gives rise to the factor $1+G(q^2)$ in \1eq{masseq}.

Note that, in the Landau gauge only,
the quantity $1+G(q^2)$ 
is linked to the inverse of the ghost dressing function $F(q^2)$
through~\cite{Binosi:2009qm,Grassi:2004yq}
\be
F^{-1}(q^2) \approx 1+G(q^2).
\label{funrelapp}
\ee
This relation, which is valid to a very good approximation, and becomes an exact equality at $q^2=0$,  
allows one to use the lattice results of~\cite{Bogolubsky:2009dc} 
for the ghost dressing function, in order to determine $G(q^2)$. 

It is obvious that 
the function $\Y(k^2)$  represents a crucial ingredient 
of \1eq{masseq}. However, its exact closed form is not available, 
mainly  because our present knowledge of the full three-gluon vertex, entering in its definition,     
is incomplete (for recent studies see~\cite{Huber:2012kd,Cucchieri:2008qm}).  
We must therefore resort to approximate expressions for this quantity.
In particular, we will employ the lowest-order perturbative expression for  $\Y(k^2)$,
obtained from \1eq{theY} by substituting the tree-level values for all quantities appearing there.  
Within this approximation, and after carrying out momentum subtraction renormalization (MOM) at $k^2=\mu^2$, one finds~\cite{Binosi:2012sj}
\be
Y_{\mathrm R}(k^2)=-\frac {\alpha_s C_A}{4\pi}\frac{15}{16}\log\frac{k^2}{\mu^2},
\label{Yappr}
\ee
where $\alpha_s$ is the value of the coupling at the subtraction point chosen.
This simple approximation will be compensated, in part,  
by multiplying $Y_{\mathrm R}(k^2)$ by and arbitrary constant $C$, 
\ie by implementing the replacement 
$Y_{\mathrm R}(k^2) \to C\, Y_{\mathrm  R}(k^2)$ and 
treating $C$ as a free parameter.
In this heuristic way, one hopes to model 
further  corrections that may  be added  to the
``skeleton'' result  provided by~\1eq{Yappr}. 
For a more sophisticated analysis based on the renormalization group properties of \1eq{masseq}, see~\cite{Aguilar:2014tka}.


\section{The unquenching formula: getting $\DeltaQ(q^2)$ from $\Delta(q^2)$}

\begin{figure}[!t]
\begin{center}
\includegraphics[scale=0.8]{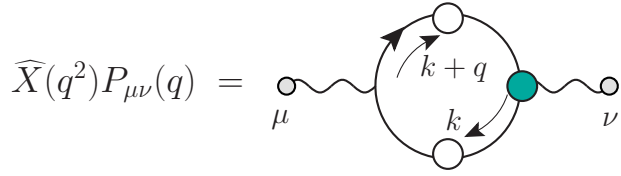}
\end{center}
\vspace{-0.5cm}
\caption{\label{Ferloop} Diagrammatic representation 
of the nonperturbative quark loop, $\widehat{X}_{\mu\nu}$,  
which determines, at this level of approximation, the unquenching effects.  
The fully-dressed (green) vertex represents the PT-BFM vertex $\widehat{\Gamma}_\mu$.} 
\end{figure}

As has been explained in~\cite{Aguilar:2012rz}, 
when the number of quark families is relatively small, 
it is reasonable to assume that the main bulk of the unquenching
effects is captured by the (fully dressed) one-loop diagram of \fig{Ferloop}
neglecting, at this level of approximation,   
all contributions stemming from (higher order) diagrams 
containing nested quark loops.

Let us now turn to the form of this particular quark-loop diagram within the PT-BFM scheme.
Factoring out the trivial color structure $\delta^{ab}$, we obtain  
\be
\quark^{\mu\nu}(q^2)=-g^2\,d_f\!\kint\mathrm{Tr}\left[
\gamma^\mu S(k)\widehat\Gamma^\nu(k+q,-k,-q)S(k+q)\right] \,,
\label{qse}
\ee
where $d_f$ is the Dynkin index of the fundamental representation [$d_f=1/2$ for $SU(3)$],
and $S$ denotes the full quark propagator, 
where, in the usual notation,
\be
S^{-1}(p)=-i\left[A(p)\pslash-B(p)\right]=-iA(p)\left[\pslash-\qm(p)\right] \,,
\label{qprop}
\ee
and the ratio \mbox{$\qm(p)=B(p)/A(p)$} is the dynamical quark mass.
The vertex $\widehat{\Gamma}_\mu$ corresponds to the PT-BFM quark-gluon vertex, satisfying the QED-like 
Ward identity~\cite{Binosi:2009qm}
\be
iq^\mu \widehat{\Gamma}_\mu(k,-k-q,q)=S^{-1}(k)-S^{-1}(k+q).
\label{GWI}
\ee

As a consequence of this Ward identity, it is immediate to show that 
\be
q^{\mu} \widehat{X}_{\mu\nu}(q) =0 \,,
\ee
a fact that allows one to cast $\widehat{X}_{\mu\nu}$ into the form  
\be
\widehat{X}_{\mu\nu}(q)=\widehat{X}(q^2)P_{\mu\nu}(q). 
\label{Drnf}
\ee 
Then, taking the trace $\quark^{\mu}_{\mu}(q)$, we obtain from \1eq{qse} the expression                       
\be
\widehat{X}(q^2)=-\frac{g^2}{6}\!\int_k\mathrm{Tr}\left[\gamma^\mu S(k)\widehat\Gamma_\mu(k,-k-q,q)S(k+q)\right].
\label{quark-loop-full}
\ee 

The simple WI satisfied by the vertex $\widehat{\Gamma}_\mu$ turns out to be particularly convenient, 
because it allows one to employ a simple Ansatz for its longitudinal part. In particular,  
one may use the standard expression known form the studies of QED, namely~\cite{Curtis:1993py}
\be
\widehat\G^\mu(p_1,p_2,p_3) =\frac{A(p_1)+A(p_2)}{2}\gamma^{\mu} 
+\frac{(p_1-p_2)^{\mu}}{p_1^2-p_2^2}\left\{\left[A(p_1)-A(p_2)\right] 
\frac{\sla{p_1}-\sla{p_2}}{2}
+\left[B(p_1)-B(p_2)\right] 
\right\} \,.
\label{bcvertex}
\ee
The main advantage of this Ansatz is that it does not involve the 
so-called quark-ghost kernel, whose general structure is only partially known. 

Of course, in the case of including various quark loops, 
corresponding to different quark flavors $\nf$,  the term $\quark^{\mu\nu}(q)$ in \1eq{Drnf}  
is replaced simply by the sum over all quark loops, \ie 
\be
\quark^{\mu\nu}(q) \to \sum_{f}\widehat{X}_{f}^{\mu\nu}(q).
\label{sumX}
\ee

Then, through the detailed analysis of~\cite{Aguilar:2012rz} one reaches the conclusion that 
the unquenched gluon propagator $\DeltaQ(q^2)$ may be expressed as a deviation from the  
quenched propagator $\Delta(q^2)$, namely (Euclidean space)
\be
\DeltaQ(q^2) = \frac{\Delta(q^2)}
{1 + \left\{ \quark(q^2) \left[1+G(q^2)\right]^{-2}+ \lambda^2(q^2)\right\}\Delta(q^2)}, 
\label{mastformeuc}
\ee
where the quantity 
\be
\lambda^2(q^2)=\mQ(q^2)-m^2(q^2) \,,
\label{lambda}
\ee 
measures the difference induced to the gluon mass due to the inclusion of quarks.
In particular, $\mQ(q^2)$ is to be obtained from \1eq{masseq} by implementing on its rhs the 
substitution \mbox{$\Delta(q^2)\to \DeltaQ(q^2)$}.

We emphasize that, as one can demonstrate using a special identity~\cite{Aguilar:2012rz},   
the nonperturbative $\widehat{X}(q^2)$ vanishes at the origin, 
$\widehat{X}(0)=0$, exactly as it happens in perturbation theory. This formal property 
is captured clearly in the numerical evaluation of $\widehat{X}(q^2)$, shown in \fig{quark-loop}.

Thus, the inclusion of quark loops 
affects the value of the saturation point of the gluon propagator not {\it directly} 
through the presence of the $\widehat{X}(q^2)$, but rather {\it indirectly} 
through the generation of a non vanishing mass difference $\lambda^2(q^2)$.

\begin{figure}[!t]
\begin{center}
\includegraphics[scale=0.7]{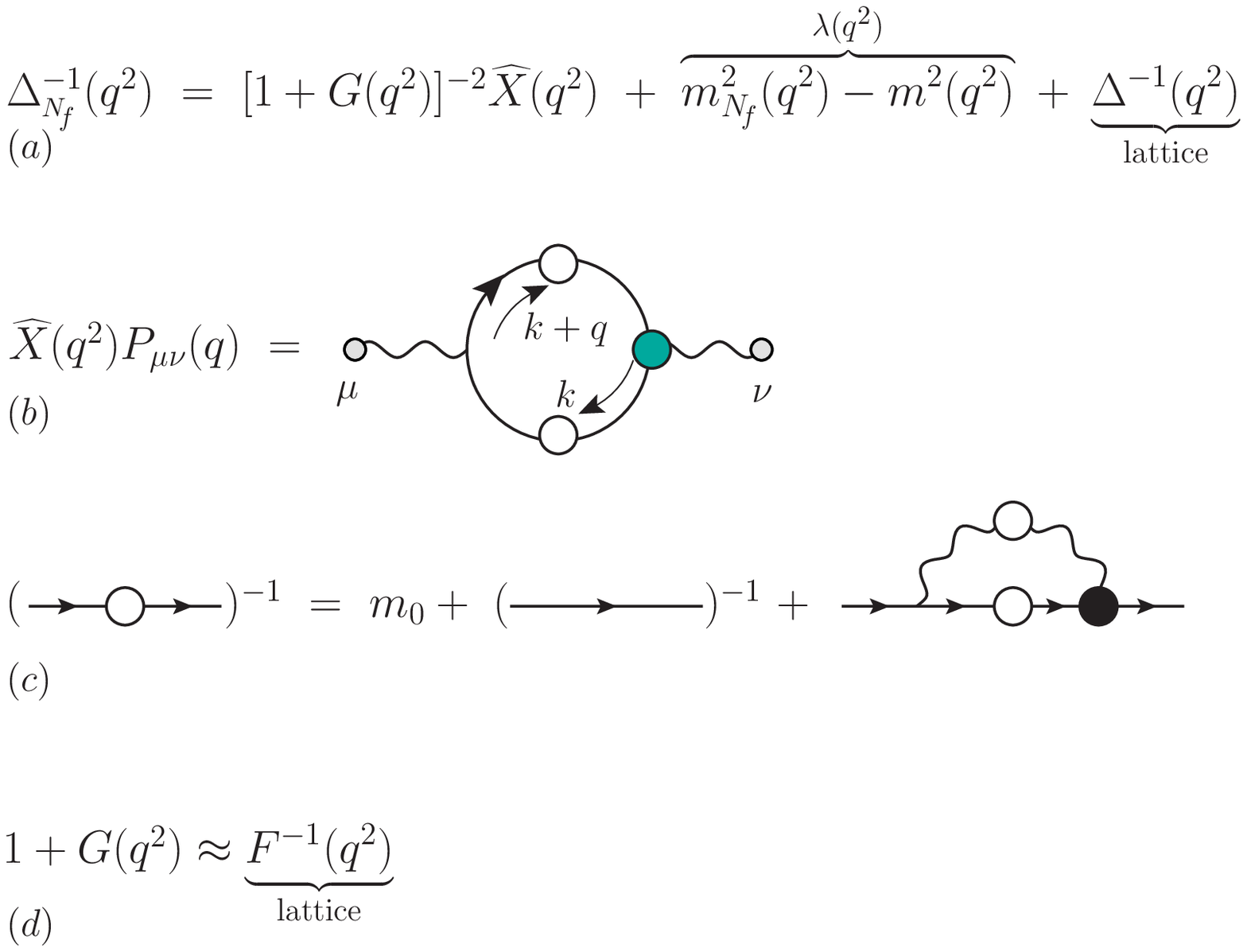}
\end{center}
\vspace{-0.5cm}
\caption{\label{Unq-gl-SDE} Schematic representation of the unquenched propagator $(a)$,
corresponding to Eq.(3.9), and some of the ingredients [$(b)$, $(c)$, and $(d)$] entering in it.
In particular, $(b)$ represents the quark loop, which, in the approximation employed, is the
only source of quark-dependence.
The quark propagators entering in $(b)$ are solutions of the gap equation
depicted in $(c)$, where $m_0$ denotes the appropriate current mass.
Finally, the function $G(q^2)$ is obtained from the relation shown in $(d)$.
The quantities obtained from the lattice are also indicated. Note that 
$\lambda^2(q^2)$ will be determined {\it dynamically}, once the
mass equation is coupled to $(a)$.} 
\end{figure}

The next step is to treat the mass equation \1eq{masseq} and the unquenching master 
formula of \1eq{mastformeuc} as a {\it coupled system}, and determine simultaneously both $\mQ$ and $\DeltaQ(q^2)$. 
In doing that, 
the nonperturbative form of each quark propagator entering into $\widehat{X}(q^2)$ 
will be obtained from the standard gap equation~\cite{Aguilar:2010cn}, supplemented by  
an appropriate current mass term, in order to make contact with the lattice results of~\cite{Ayala:2012pb}.
In this latter simulation,   
the gluon (and ghost) propagators have been evaluated from large volume configurations (up to $3^3\times6$ [fm$^4$]),  
generated from a lattice action that included (twisted mass) fermions. 
Specifically, one employed two  light degenerate quarks ($\nf=2$),  
with a current mass ranging from 20 to 50 MeV, or two light and two heavy quarks ($\nf=2+1+1$), 
with a strange (charm) quark current mass  roughly set to 95 MeV (\mbox {1.51 GeV}). 
Thus, effectively, one ends up dealing with the rather extended set of equations depicted in~\fig{Unq-gl-SDE}. 

\section{Numerical results} 

\begin{figure}[!t]
\begin{minipage}[b]{0.45\linewidth}
\centering
\includegraphics[scale=0.50]{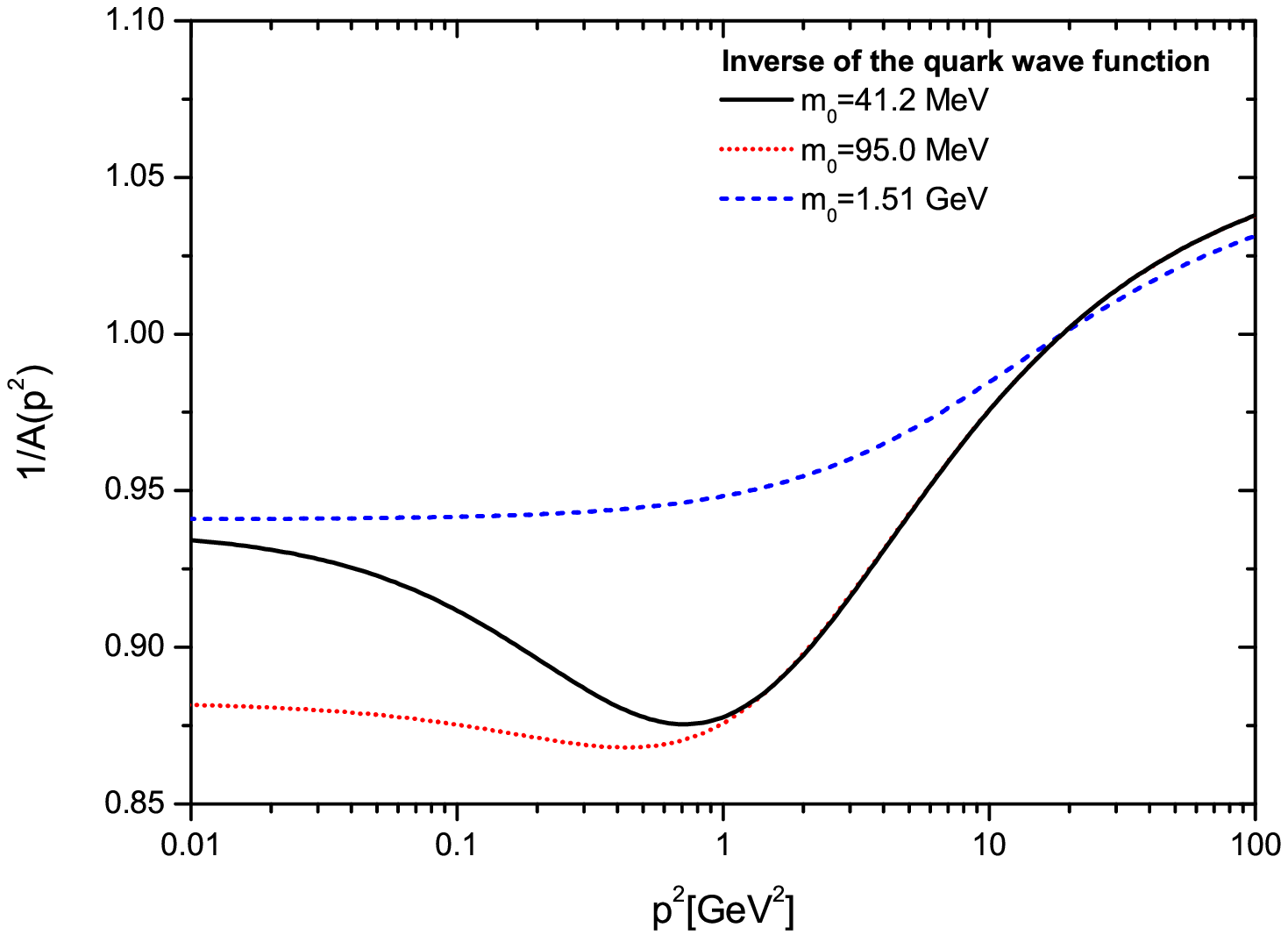}
\end{minipage}
\hspace{0.5cm}
\begin{minipage}[b]{0.50\linewidth}
\includegraphics[scale=0.50]{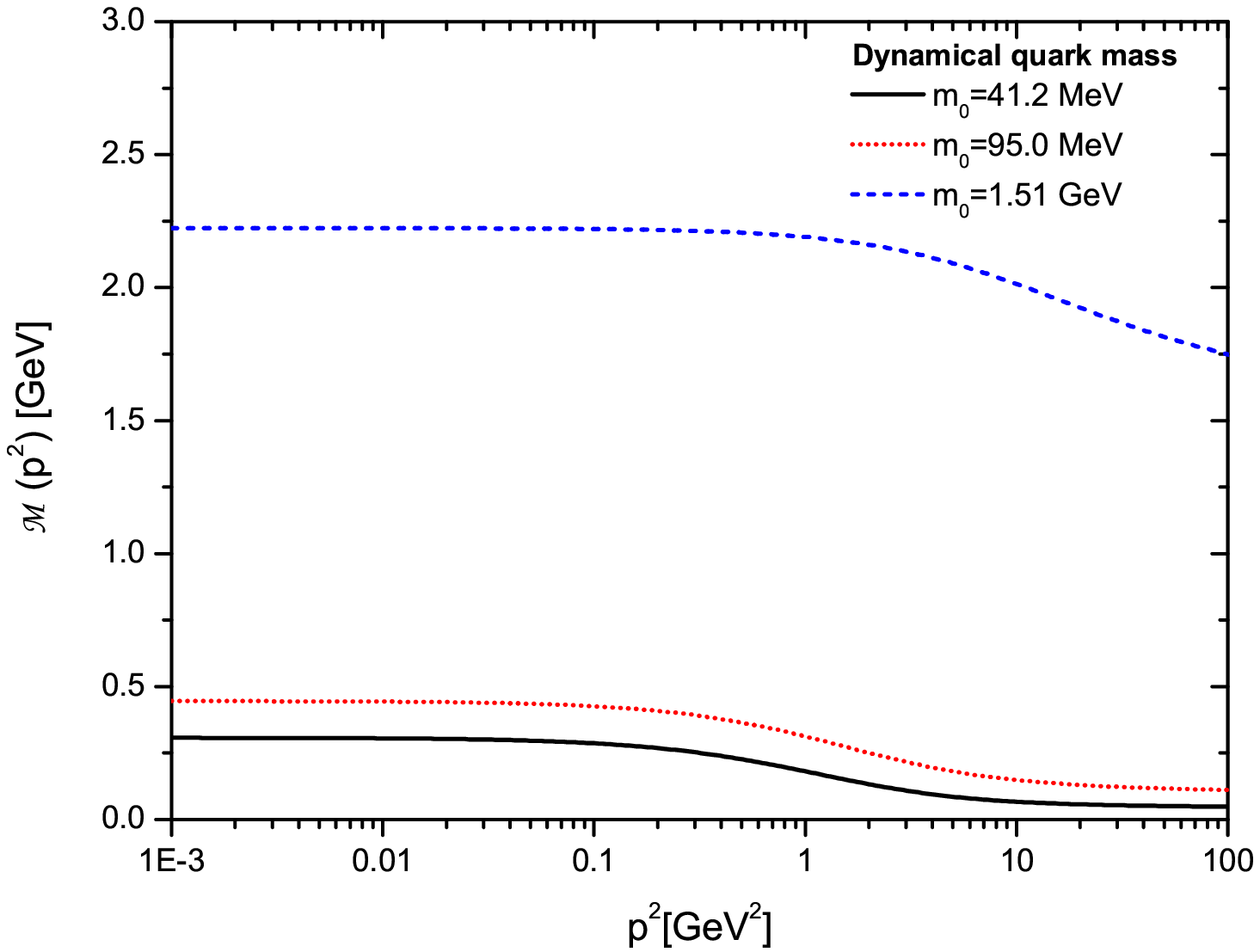}
\end{minipage}
\caption{ 
The inverse of the quark wave-functions (left panel), and the dynamical  quark masses (right panel), obtained from 
the quark gap equation for three different values of the 
current mass: $m_0=41.2$ MeV (black, continuous), $m_0=95$ MeV (red, dotted) and $m_0=1.51$ GeV (blue, dashed). }
\label{CSBquark-mass_wave}
\end{figure}

The system of SDEs that we consider is composed 
of~\3eqs{masseq}{quark-loop-full}{mastformeuc}, supplemented by the
quark  gap equation.  The initial  condition is provided  by the
quenched $SU(3)$ gluon propagator and ghost dressing function obtained
in the lattice  simulations of~\cite{Bogolubsky:2009dc}, which will be
also used to  determine the initial values of  the form factors $A(p)$
and  ${\cal M}(p)$. All  calculations will  be performed using propagators  
renormalized at $\mu=4.3$ GeV.

The algorithm that we employ consists of  the following main steps.

\n{i} We use the  quenched propagator  as an input  of the  first iterative
step, and determine the quark form factors $A(p)$ and ${\cal M}(p)$ by
solving  the quark  gap  equation (see the results in Fig.~\ref{CSBquark-mass_wave}).

\begin{figure}[!t]
\begin{center}
\includegraphics[scale=1]{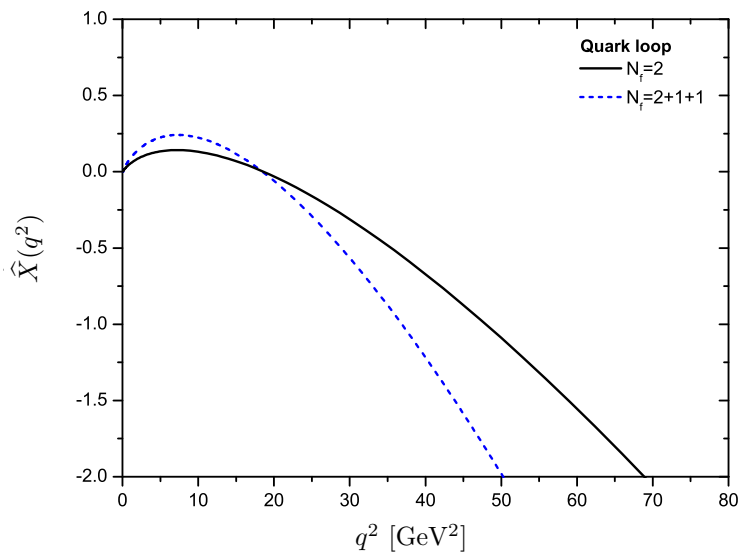}
\end{center}
\vspace{-0.5cm}
\caption{\label{quark-loop} The full nonperturbative quark loop contribution $\widehat{X}(q^2)$ for the two cases $\nf=2$ (black, continuous) and $\nf=2+1+1$ (blue, dashed).}
\end{figure}

\n{ii} $A(p)$ and ${\cal M}(p)$ are then 
substituted  into~\1eq{quark-loop-full}, and  the corresponding  value of
the  quark loop diagram  $\widehat{X}(q^2)$ is evaluated (results in Fig.~\ref{quark-loop}).

\n{iii} The preliminary form of $\DeltaQ(q^2)$  is determined from~\noeq{mastformeuc},              
employing            initially
$\lambda^2(q^2)=0$, with the  quenched mass $m^2(q^2)$ obtained
from the  solution of the  mass equation~\noeq{masseq} corresponding
to  the  quenched   lattice  propagator.

\n{iv} The unquenched propagator 
$\DeltaQ(q^2)$ of the previous step   is substituted    into   the   mass
equation~\noeq{masseq}   in  order   to  determine   the  associated
unquenched dynamical  gluon mass $\mQ(q^2)$,  and therefore the
corresponding $\lambda^2(q^2)$. 

\n{v} At this point the  latter quantity is
inserted back  into the master  equation~\noeq{mastformeuc}, and the
loop starts again, until convergence, determined by the 
stability of the quantities involved, has been reached.

\begin{figure}[!t]
\hspace{-1.0cm}
\begin{minipage}[b]{0.45\linewidth}
\centering
\includegraphics[scale=0.65]{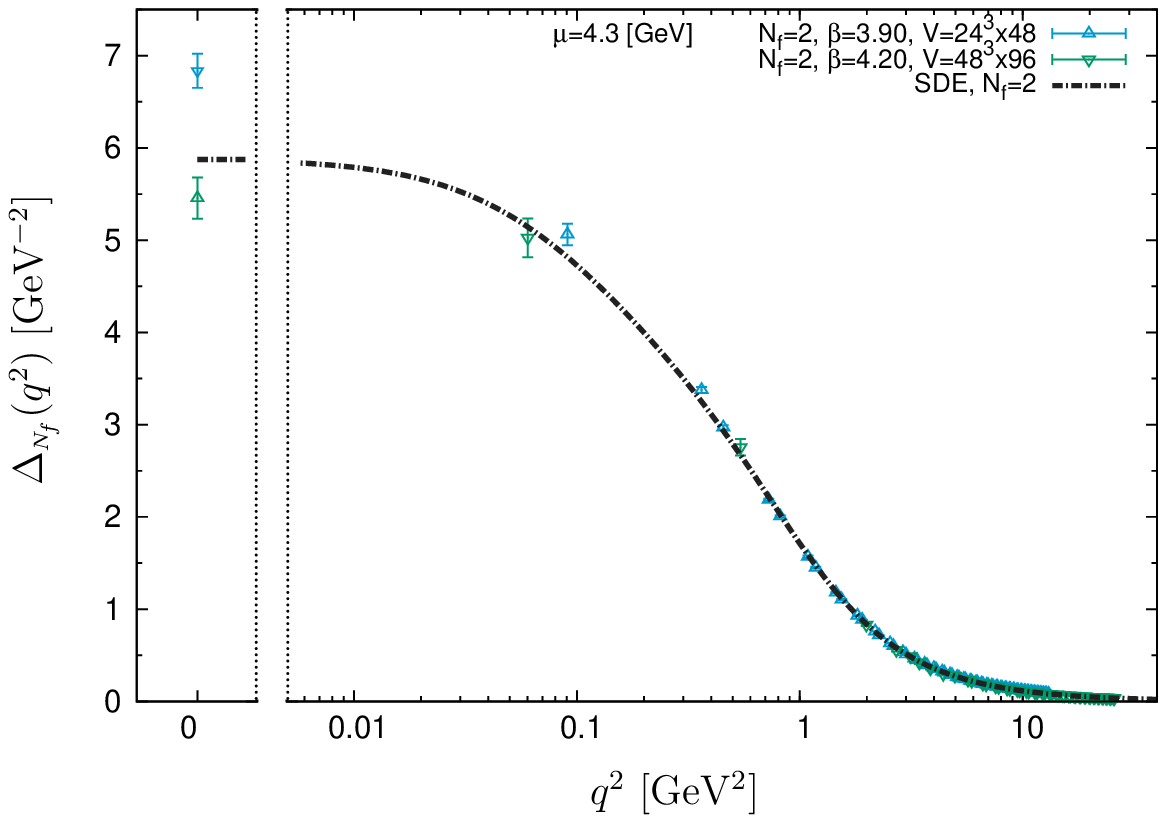}
\end{minipage}
\hspace{2.0cm}
\begin{minipage}[b]{0.50\linewidth}
\hspace{-1.0cm}
\includegraphics[scale=0.65]{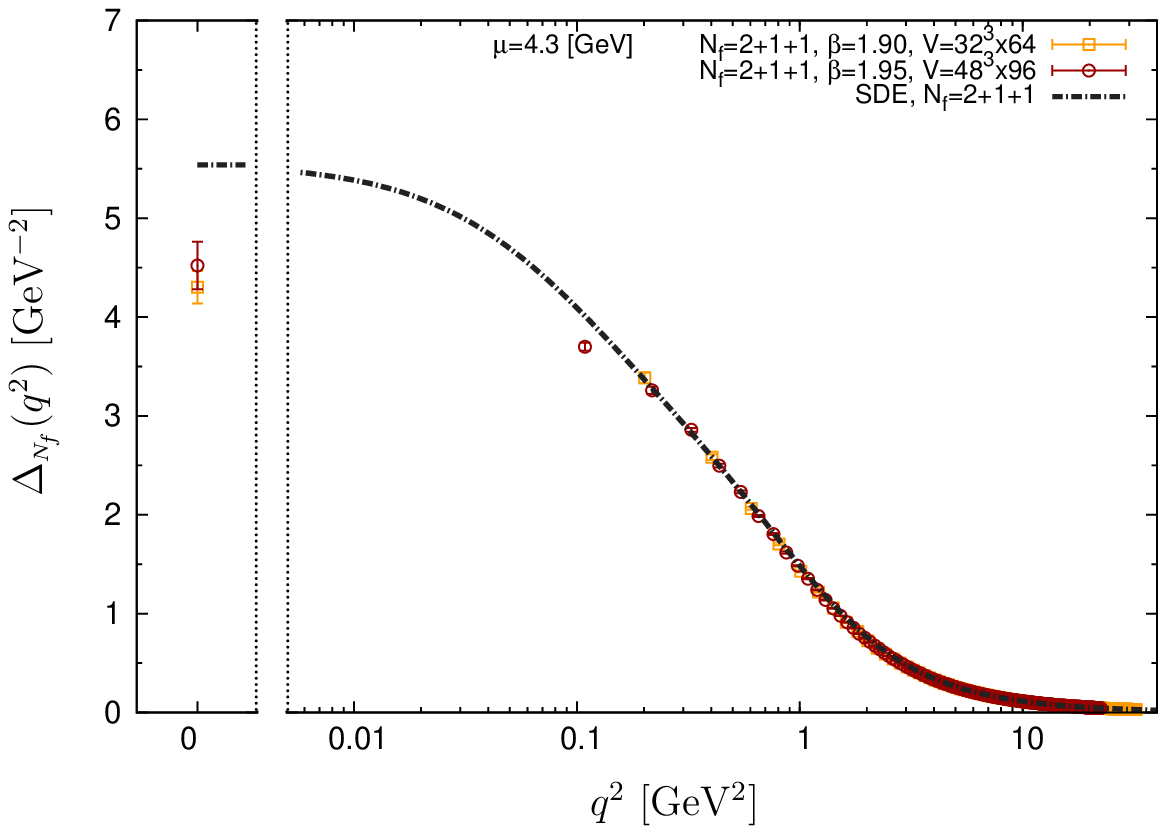}
\end{minipage}
\vspace{-0.7cm}
\caption{\label{res_fig} The unquenched gluon propagators obtained from our analysis, 
for  $\nf=2$ (left panel) and $\nf=2+1+1$ (right panel), compared with the lattice data of~\cite{Ayala:2012pb} for the same cases.}
\end{figure}

In~\fig{res_fig} we present the central result of this analysis. In particular,  
we plot the  propagators obtained  
when convergence of the above mentioned iteration procedure has been reached,  
and compare them with  the  corresponding  unquenched  lattice data,  recently  reported
in~\cite{Ayala:2012pb}.   We  observe   a rather good agreement
between our  theoretical predictions  and the lattice  computation for
both  values  of  $\nf$, for the available range of physical momenta. 
A notable exception to this fair coincidence between curves is the   
saturation  point  of the $\nf=2+1+1$ case; specifically, 
the value obtained from our SDE analysis is  20\% higher than that  found 
in lattice simulations. 

\begin{figure}[!t]
\hspace{-0.5cm}
\begin{minipage}[b]{0.45\linewidth}
\centering
\includegraphics[scale=0.59]{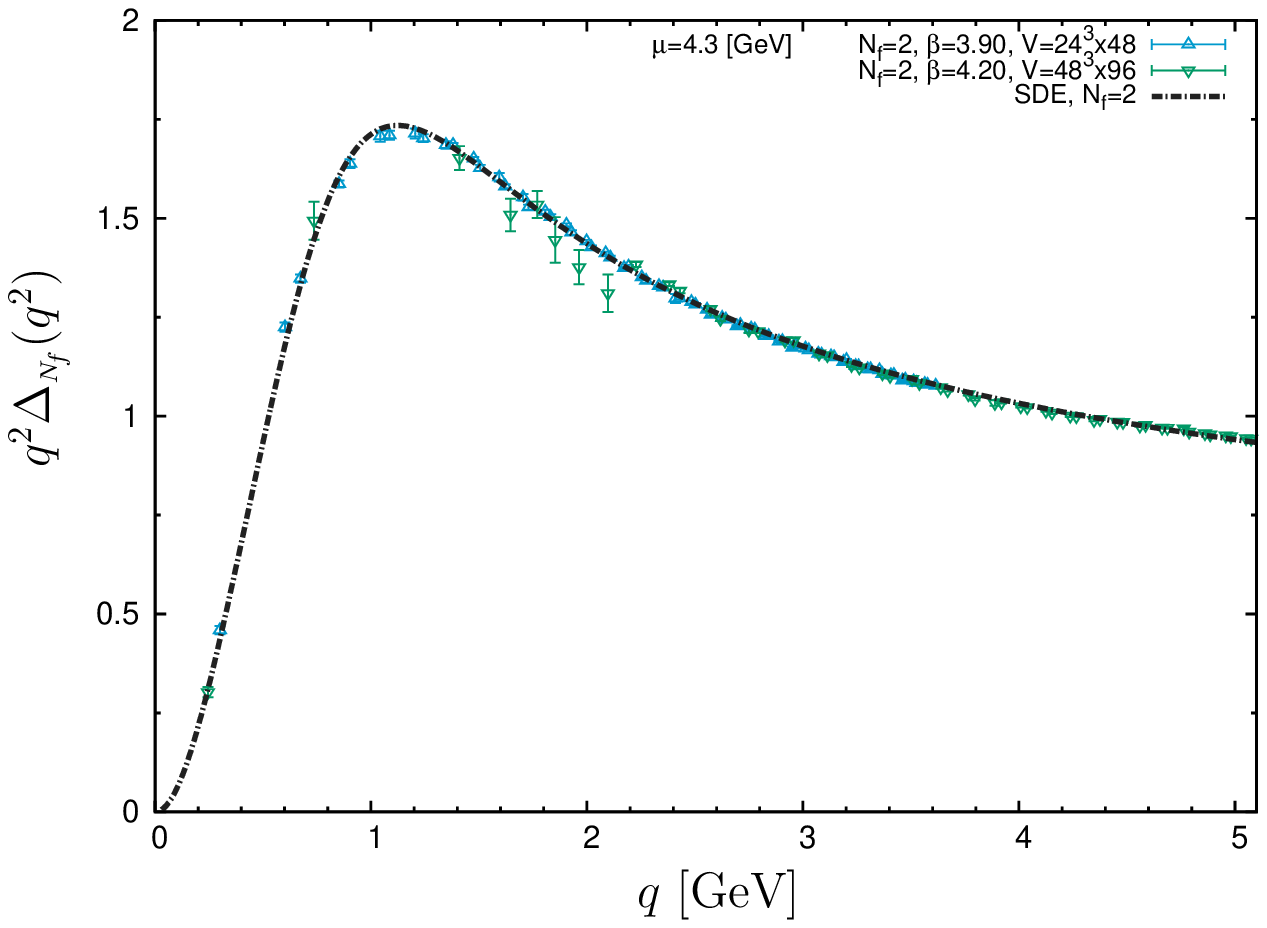}
\end{minipage}
\begin{minipage}[b]{0.50\linewidth}
\hspace{0.9cm}
\includegraphics[scale=0.59]{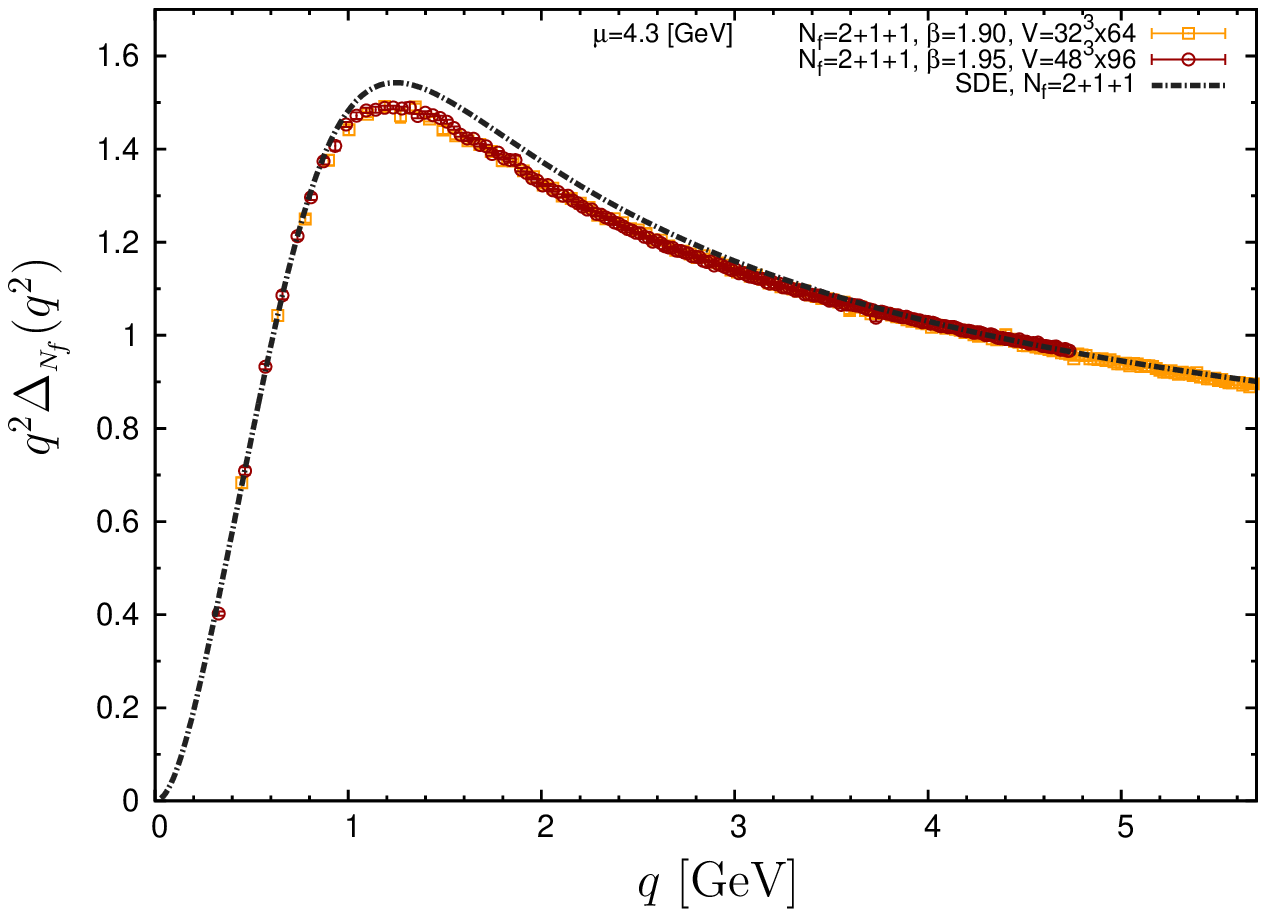}
\end{minipage}
\vspace{-0.2cm}
\caption{\label{res_dr_fig} The unquenched gluon dressing functions obtained from our analysis, 
for  $\nf=2$ (left panel) and $\nf=2+1+1$ (right panel), compared with the lattice data of~\cite{Ayala:2012pb} for the same cases.}
\end{figure}

Similar conclusions can be drawn by observing the 
plot corresponding to the gluon dressing functions, $q^2\DeltaQ(q^2)$, shown in \fig{res_dr_fig}: 
while one has an excellent agreement in the case of two degenerate light quarks, when two heavier quarks are added the SDE solution tends to mildly 
overestimate the amplitude of the characteristic peak,x located in the intermediate  momentum region.

\begin{figure}[!t]
\begin{center}
\includegraphics[scale=1]{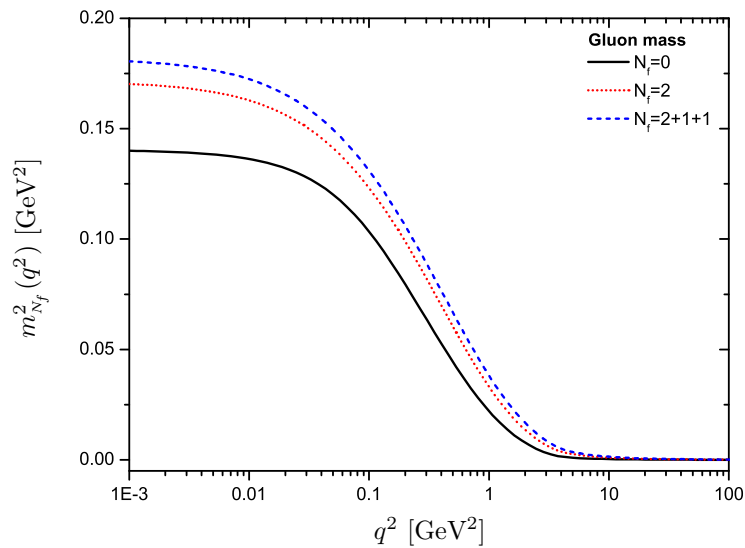} 
\end{center}
\vspace{-0.7cm}
\caption{\label{gluonmass_fig} Solution of the mass equation yielding the dynamically generated gluon mass for $\nf=2$ (red dotted line) and $\nf=2+1+1$ (blue dashed line). In the deep infrared one has \mbox{$m_2(0)=413$ MeV}, and \mbox{$m_{2+1+1}(0)=425$ MeV}. For comparison we also show the quenched gluon mass (black continuous line) obtained from the quenched lattice propagator, in which case \mbox{$m(0)=376$ MeV}.} 
\end{figure}

The corresponding dynamical gluon masses,  $m^2_2(q^2)$ and $m^2_{2+1+1}(q^2)$, are shown in~\fig{gluonmass_fig}; 
for comparison,  we also plot the quenched solution, $m^2(q^2)$, obtained  
from \1eq{masseq} when the quenched lattice propagators of~\cite{Bogolubsky:2009dc} are used as input. 
In particular, the corresponding 
saturation points give $m_2(0)=413$ MeV and $m_{2+1+1}(0)=425$ MeV (at $\mu=4.3$ GeV), 
which should be compared with the value $m(0)=376$ MeV found in the quenched case.
The results captured in ~\fig{gluonmass_fig} are particularly important, because they demonstrate clearly 
that the mass generation mechanism established for pure Yang-Mills 
continues to operate in QCD-like circumstances.

A this point it seems reasonable  to think that the
observed deviation between our results and the lattice 
signals a mild  violation of one of the assumptions
underlying      the       derivation      of      the      unquenching
formula~\noeq{mastformeuc}. 
In  particular, it is natural to expect that our main operating hypothesis, namely that  
the quark-loop contributions constitute a ``perturbation''  of   
the   quenched propagator, becomes progressively less 
accurate as the number of active flavors increases. 
It is therefore possible that from $\nf>2$ onward we begin to perceive 
the onset of additional effects, not captured by~\noeq{mastformeuc}.

In particular, 
the ``lowest order unquenching''  assumed here 
includes explicitly only the contribution of the quark loop $\widehat{X}(q^2)$, keeping all other quantities unquenched. 
This is reflected clearly at the level of the master formula \1eq{mastformeuc}, where the quantity $1+G(q^2)$ 
(or, equivalently, $F^{-1}(q^2)$, by virtue of \1eq{funrelapp}) 
assumes its quenched form, obtained from \cite{Bogolubsky:2009dc}.  
Moreover, the computation of $\widehat{X}(q^2)$  [see \fig{Unq-gl-SDE}] uses as input the  
quark propagator obtained from the gap equation, which, in turn, depends on both the gluon propagator 
and the ghost dressing function; again, the quenched forms of \cite{Bogolubsky:2009dc} were employed.
Finally, the strength of the gauge coupling $g$ also depends on the number of flavors; in the present analysis 
we have used its value when $\nf=0$ (MOM~\cite{Boucaud:2008gn}).
In order to improve this analysis, and 
eventually reach a better agreement with the lattice, one could  
gradually introduce quark effects into some of 
the aforementioned (quenched) ingredients.
For example, one could envisage the possibility of using unquenched instead of quenched data for the ghost dressing function 
$F(q^2)$, obtained from the lattice analysis of~\cite{Ayala:2012pb}. Given that this quantity enters both 
in the master formula and the gap equation, its overall effect may be appreciable. 
In addition, the increase 
in the value of the gauge coupling produced by the inclusion of quark flavors may modify our predictions in the 
direction of the lattice data.

Furthermore, an additional theoretical uncertainty originates from the 
approximate (perturbative) treatment of the quantity $Y(k^2)$. In particular, the parameter $C$  
may only model, to some extent, unknown contributions that display 
a logarithmic momentum dependence, 
as in \1eq{Yappr}, but cannot account for terms with a different functional form. 
Moreover, the use of an Ansatz for the vertex $\widehat\Gamma_{\mu}$  
entering into the definition of ${\widehat X}(q^2)$ may 
induce further error, due to the fact that its 
transverse (automatically conserved part) is in general undetermined.


\section{Conclusions}

In this presentation, by  employing a methodology  relying mainly on  the SDEs
that describe the gluon two-point sector within the PT-BFM  framework, 
we  studied in quantitative detail how the 
inclusion of dynamical quarks
affects  the generation  of  the momentum-dependent  gluon mass, in the 
Landau gauge.  
Our main conclusion is that the 
gluon  propagator continues to 
saturate in the infrared, due to the dynamical generation of a gluon mass.
In fact, the analysis suggests that the gluon mass becomes heavier 
as the number of active quark families increases. 
It would be interesting to study the possible limitations of this picture,
and determine whether there is a critical number of quark families, 
past which the gluon generation mechanism clashes with the quark-induced 
dynamics~\cite{Cheng:2011qc,Tomboulis:2012nr} .

\vspace{0.5cm}

{\it Acknowledgements:} 

\vspace{0.5cm} 

I would like to thank the ECT* for making this workshop possible.
This research is supported by the Spanish MEYC under 
grant FPA2011-23596.

\end{document}